\documentclass[12pt]{article}
\usepackage{amsmath,amssymb}
\usepackage{graphicx}
\usepackage{color}
\usepackage{url}
\usepackage{multirow}

\begin{document}
\baselineskip=16pt
\begin{titlepage}
\begin{flushright}
{\small EPHOU-12-009}\\
{\small OU-HET 769/2012}\\
{\small KEK Preprint 2012-30}\\
\end{flushright}
\vspace*{1.2cm}

\begin{center}

{\Large\bf 
QCD parity violation at LHC in 
warped extra dimension 
} 
\lineskip .75em
\vskip 1.5cm

\normalsize
{\large Naoyuki Haba}$^1$,
{\large Kunio Kaneta}$^{1,2}$, and
{\large Soshi Tsuno}$^3$

\vspace{1cm}
$^1${\it Department of Physics, 
Faculty of Science, Hokkaido University, Sapporo 060-0810, Japan}\\
$^2${\it Department of Physics, 
 Osaka University, Toyonaka, Osaka 560-0043, 
 Japan} \\
$^3${\it High Energy Accelerator Research Organization (KEK), 
Tsukuba, Ibaraki 305-0801, Japan}

\vspace*{10mm}

{\bf Abstract}\\[5mm]

{\parbox{13cm}{\hspace{5mm}

Extra dimension is one of the most attractive candidates
 beyond
 the Standard Model.
In warped extra dimensional space-time, 
 not only gauge hierarchy problem but also 
 quark-lepton mass hierarchy can be naturally explained. 
In this setup,
 a sizable parity violation 
 through 
 Kaluza-Klein gluon exchange   
 appears in QCD process  
 such as helicity dependent top pair production. 
We investigate this QCD parity violating process  
 by use of $SO(5)\times U(1)$ gauge-Higgs unification model. 
We evaluate LHC observable quantities, i.e., 
 a charge asymmetry and a forward-backward asymmetry
 of the top pair production, 
 and find that 
 a sizable charge asymmetry
 can be observed 
 with specific model parameters.

%

}}

\end{center}

\end{titlepage}
\section{Introduction}

ATLAS and CMS collaborations at the LHC reported
 a discovery of new boson 
 which is consistent with the Standard Model (SM)
 Higgs boson \cite{CMS-ATLAS}.
It is important that
 the observed boson can be really identified to the SM Higgs boson. 
On the other hand, 
 a stabilization of Higgs self-coupling 
 requires underlying theory behind the SM 
 \cite{Cabibbo:1979ay,EliasMiro:2011aa}
 (see also \cite{Djouadi:2005gi} and references therein).
Supersymmetry (SUSY) and extra dimension are 
 most reliable 
 candidates beyond the SM, 
 which naturally contain stable dark matter particles.  
%
For 
 warped extra dimension,
 which is first proposed by 
 Randall and Sundrum (RS) in
Ref.\cite{Randall:1999ee}, provides a framework which solves the hierarchy problem.
In the original model, the SM fields are localized to a brane.
However, when the SM fermions and gauge bosons propagate in the bulk,
models have attractive features such as explaining fermion mass
hierarchy 
 (see, for example \cite{Grossman:1999ra}).
In this setup, 
 configuration of the SM fermion wave function depends on bulk mass parameters $c_i$, 
where $i$ is a label of fermion.
Fermions with $c_i > 1/2$ are localized near the Planck brane,
while the ones with $c_i < 1/2$ are localized near the TeV brane.
Since the Higgs
 is localized to the TeV brane,
mass of fermions with $c_i > 1/2$ is smaller than that of fermions with $c_i < 1/2$
due to overlap of wave functions among the Higgs and fermions.
In general, $c_i$ of left-handed fermions are not the same as those of right-handed fermions\cite{Huber:2003tu}.
Focusing attention on QCD sector,
 $n$th Kaluza-Klein (KK) gluon $G^{(n)}$ is localized to the TeV brane.
Therefore overlap between $G^{(n)}$ and $q_L$
 is different from that between $G^{(n)}$ and $q_R$\cite{Lillie:2007yh}.
This means that parity violation in QCD process
 is accommodated in warped extra dimension scenario.
Parity violation in QCD process can be measured by using helicity dependent top pair production.
Helicity measurement of $t\bar t$
is shown in Ref.\cite{Stelzer:1995gc}. 
In the SM QCD sector, 
of course,
 there is no parity violation in top pair production. 
The SM background is induced by electroweak
 interaction\cite{Beenakker:1993yr,Haba:2011cj}.
The $t \bar t$ helicity asymmetry is expected to be highly sensitive to new physics.
For example, SUSY can also arise
 sizable $t \bar t$ asymmetry through
 squark loop diagrams, 
 because 
 $\tilde q_L$ and $\tilde q_R$ have different mass spectrum in general,
 and $q_{L(R)}$-$\tilde q_{L(R)}$-$\tilde g$ is
 chiral interaction\cite{Haba:2011cj}.
For another QCD parity violating process, 
 quarkonium decay is investigated in Ref. \cite{Haba:2011ib}.
Comparing to the SUSY models, 
 the warped extra dimension model 
 has much larger
 QCD parity violation due
 to an existence of tree level contributions.

In this paper,
 we investigate the QCD parity violation 
 by use of $SO(5)\times U(1)$  
 gauge-Higgs unification model as an example 
 of warped extra dimension scenario 
 with bulk quark configurations. 
That is, 
 Higgs and $G^{(n)}$ are localized to the TeV-brane,
 and $q_L (q_R)$ is typically located near
 the Planck (the TeV) brane\cite{Hosotani:2011vr}.\footnote{
In general, configuration of quark wave function is also different among their flavor.
Thus a different configuration between $q_L$ and $q_R$ induces not only parity violation but also flavor violation.
Constraints from flavor violation are studied in
Refs.\cite{Blanke:2008zb}.}
We investigate 
 helicity asymmetry of top pair production. 
It was also researched in 
 Refs.\cite{Lillie:2007yh},  
 however, we 
 will evaluate it by use of 
 LHC observables, i.e., 
 a charge asymmetry and a
 forward-backward asymmetry here. 
We will find
 that a sizable charge asymmetry can be observed 
 with specific model parameters. 

This paper is organized as follows.
In section 2,
we give a brief review of $SO(5)\times U(1)$ gauge-Higgs unification model.
In section 3,
 we analyze the $t\bar t$ 
 left-right asymmetry $A_{LR}$, 
 which can be observed by a 
 charge asymmetry $A_C$, and a forward-backward asymmetry $A_{FB}$ 
 in the LHC experiment.
We present a conclusion in section 4.

\section{$SO(5)\times U(1)$ gauge-Higgs unification model}

We pick up $SO(5)\times U(1)$ gauge-Higgs unification model 
 as a warped extra dimension scenario in which 
 Higgs and KK gluon are localized to the TeV brane 
 and left- and right-handed fermions have
 different configurations in the bulk. 
The model is constructed in the RS 
 warped space-time \cite{Randall:1999ee}, 
 which metric is given by
\begin{equation}
ds^2 \; = \; e^{-2\sigma(y)}\eta_{\mu\nu}dx^\mu dx^\nu + dy^2,
\end{equation}
where $\sigma(y)=k|y|$ with 5-dimensional scalar curvature $k$.
The fifth dimension $y$ is orbifolded on $S^1/Z_2$, and the region of $y$ is given by $0 \leq y  \leq L$.
The Planck and the TeV brane are located at $y = 0$ and $y = L$, respectively.
Gauge group of this model is $SO(5)\times U(1)_X\times SU(3)_C$ in the bulk,
and $SO(5)\times U(1)_X$ is reduced to $SU(2)_L\times SU(2)_R\times U(1)_X$ by orbifold boundary conditions.
The $SU(2)_R\times U(1)_X$ symmetry breaks down to $U(1)_Y$ by vacuum expectation value (VEV) of a scalar field $\Phi$ on the Planck brane.

At low energy scale, 
relevant parameters in QCD sector of this model are
$k$, the warp factor $z_L = e^{kL}$, 5D strong gauge coupling $g_C$, bulk mass parameters $c_q$, and brane mass ratios $\tilde \mu^q/\mu^q_2$.
The $g_C$ is related to 4D strong gauge coupling as $g_s = g_C/\sqrt{L}$.
Basically $c_q$ controls the localization of the zero mode wave functions near the TeV brane and the Planck brane.
$\tilde \mu^q$ and $\mu^q_2$ are induced by VEV of the scalar filed $\Phi$.
Brane mass matrices are regarded as flavor diagonal so that the flavor mixing is turned off in this paper.
Our setup follows in Ref. \cite{Hosotani:2011vr},
and unknown parameters are $(k, z_L, c_q, \tilde \mu^q/\mu^q_2)$ at low energy.
Three of these parameters can be fixed by taking electroweak coupling $\alpha_W$, $W$ boson mass $m_W$, and quark mass $m_q$.
One parameter of $(k, z_L, c_q, \tilde \mu^q/\mu^q_2)$ remains as a free parameter,
and we take $z_L$ as an input parameter.
In this paper we consider two cases of $z_L=10^{15}$ and $z_L=10^{20}$.
Once the $z_L$ parameter is fixed,
Higgs mass is determined.
When we input $z_L$ as $z_L=10^{15}$ and $10^{20}$,
Higgs mass is calculated
 as $m_H = 135$ GeV and 158 GeV when
 $\theta_H = \frac{\pi}{2}$, respectively.
This is not compatible with recent experimental data,
 and the suitable Higgs mass can be
 obtained when $\theta_H \neq \frac{\pi}{2}$.
Such $\theta_H$ might be realized by taking specific matter content, for example,
and QCD sector is not affected by such modification of the model.
We focus on the QCD sector of this model, and therefore, 
our analysis is not conflicted with the observed $m_H$.

In the $SO(5)\times U(1)$ gauge-Higgs unification model,
parity is violated in QCD process because of
 difference between $q_L$-$\bar{q}_L$-$G^{(n)}$ and $q_R$-$\bar{q}_R$-$G^{(n)}$ couplings.
As we show in section 3,
the latter coupling is much larger than the former one.
This is because that $q_R$ ($q_L$) wave function is located near the TeV (the Planck) brane,
and KK gluon is located near the TeV brane.
Such configuration of quark wave function is related to brane mass parameters induced by VEV of $\Phi$ on the Planck brane.
Only left-handed quark has brane mass terms,
and mixes with extra particles located on the Planck brane.
 These chiral interactions induce a sizable parity
 violation in QCD process,
 which can be discovered at LHC.
In order to investigate this parity violation,
 we focus on
 helicity dependence of top pair production at LHC.

\section{Experimental observables at LHC}
\subsection{Production cross sections}

Firstly we prepare parameter sets of the model.
Parity violation 
in QCD process 
arises from one KK gluon exchange at tree level, 
which process is $q \bar q \to G^{(n)} \to t \bar t$
. 
KK gluon masses and their total decay widths are shown in Tables \ref{KKmass} (a) with $z_L = 10^{15}$ and (b) with $z_L = 10^{20}$.
The couplings of 
KK gluon to quarks are listed in Table \ref{couplings},
where 
$g^{G^{(n)}}_{q}$ represents
$q$-$\bar q$-$G^{(n)}$ coupling constants in unit $g_s = g_C/\sqrt{L}$.
$c_q$ and $\tilde{\mu}^q/\mu^q_2$ are given in Table \ref{masspara}.

\begin{table}[htbp]
\begin{center}
\begin{tabular}{c|ccc} \hline \hline
unit GeV           & 1st KK gluon & 2nd KK gluon & 3rd KK gluon \\ \hline
mass & 1144 & 2630  & 4111  \\
$\Gamma_{\rm total}$ & 7205 & 1265 & 274.3 \\ \hline \hline
\end{tabular}
\newline
\\
(a)
\end{center}

\begin{center}
\begin{tabular}{c|ccc} \hline \hline
unit GeV           & 1st KK gluon & 2nd KK gluon & 3rd KK gluon \\ \hline
mass & 1330 & 3030 & 6452 \\
$\Gamma_{\rm total}$ & 10987 & 1615 & 175.7 \\ \hline \hline
\end{tabular}
\newline
\\
(b)
\end{center}
\caption{KK gluon masses and their total decay widths, $\Gamma_{\rm total}$, with (a) $z_L=10^{15}$ and (b) $z_L=10^{20}$.}
\label{KKmass}
\end{table}

\begin{table}[htbp]
\begin{minipage}{.5\textwidth}
\begin{center}
\begin{tabular}{c|c|c|c}
\hline\hline
unit $g_s$&$n=1$&$n=2$&$n=3$\\
\hline
$g^{G^{(n)}}_{u_L}$& -0.195 & 0.133 & -0.108 \\
$g^{G^{(n)}}_{c_L}$& -0.195 & 0.133 & -0.108 \\
$g^{G^{(n)}}_{t_L}$& 0.442 & -0.402 & 0.295 \\
$g^{G^{(n)}}_{d_L}$& -0.195 & 0.133 & -0.108 \\
$g^{G^{(n)}}_{s_L}$& -0.195 & 0.133 & -0.108 \\
$g^{G^{(n)}}_{b_L}$& 0.661 & -0.370 & 0.283 \\
&&&\\
$g^{G^{(n)}}_{u_R}$& 6.323 & 2.129 & 0.734 \\
$g^{G^{(n)}}_{c_R}$& 6.044 & 1.633 & 0.568 \\
$g^{G^{(n)}}_{t_R}$& 5.603 & 0.949 & 0.408 \\
$g^{G^{(n)}}_{d_R}$& 6.323 & 2.129 & 0.734 \\
$g^{G^{(n)}}_{s_R}$& 6.044 & 1.633 & 0.568 \\
$g^{G^{(n)}}_{b_R}$& 5.500 & 0.832 & 0.417 \\
\hline\hline
\end{tabular}
\newline
\\
(a)
\end{center}
\end{minipage}
\hfill
\begin{minipage}{.5\textwidth}
\begin{center}
\begin{tabular}{c|c|c|c}
\hline\hline
unit $g_s$&$n=1$&$n=2$&$n=3$\\
\hline
$g^{G^{(n)}}_{u_L}$&-0.168&0.114&0.079\\
$g^{G^{(n)}}_{c_L}$&-0.168&0.114&0.079\\
$g^{G^{(n)}}_{t_L}$&0.366&-0.367&-0.221\\
$g^{G^{(n)}}_{d_L}$&-0.168&0.114&0.079\\
$g^{G^{(n)}}_{s_L}$&-0.168&0.114&0.079\\
$g^{G^{(n)}}_{b_L}$&0.563&-0.334&-0.213\\
&&&\\
$g^{G^{(n)}}_{u_R}$&7.158&2.174&0.455\\
$g^{G^{(n)}}_{c_R}$&6.900&1.733&0.369\\
$g^{G^{(n)}}_{t_R}$&6.518&1.143&0.250\\
$g^{G^{(n)}}_{d_R}$&7.158&2.174&0.455\\
$g^{G^{(n)}}_{s_R}$&6.900&1.733&0.369\\
$g^{G^{(n)}}_{b_R}$&6.430&1.039&0.234\\
\hline\hline
\end{tabular}
\newline
\\
(b)
\end{center}
\end{minipage}
\caption{The coupling constants of 
$q$-$\bar q$-$G^{(n)}$ with (a) $z_L = 10^{15}$ and (b) $z_L = 10^{20}$ in unit $g_s$.}
\label{couplings}
\end{table}

\begin{table}[htbp]
\begin{center}
\begin{tabular}{c|c|c|c|c|c|c}
\hline\hline
&\multicolumn{3}{|c|}{$c_q$}&\multicolumn{3}{|c}{$\tilde{\mu}^q/\mu^q_2$}\\
\hline
$z_L$& $(u, d)$ & $(c, s)$ & $(t, b)$ & $(u, d)$ & $(c, s)$ & $(t, b)$ \\
\hline
$10^{15}$ & 0.843 & 0.679 & 0.432 & 2.283 & 0.0889 & 0.0173 \\
$10^{20}$ & 0.757 & 0.634 & 0.451 & 2.283 & 0.0889 & 0.0172 \\
\hline\hline
\end{tabular}
\end{center}
\caption{Bulk mass parameters $c_q$ and brane mass ratios $\tilde{\mu}^q/\mu^q_2$ with $z_L = 10^{15}$ and $z_L = 10^{20}$.}
\label{masspara}
\end{table}


The production cross sections of the first, second and third KK gluons 
 with the parameters of $z_{L}$=10$^{15}$ and 10$^{20}$ are summarized 
in Table \ref{topxsec}, where the top quark mass is 172.5 GeV and CTEQ6L PDF \cite{Pumplin:2002vw} is used for 
proton-proton collision at $\sqrt{s}$ = 8 TeV. For simplicity, the renormalization and 
factorization scales are fixed at $m_{Z}$ = 90.188 GeV, which result in the electroweak and 
strong coupling constants of $\alpha(m_{Z})$ = 1/132.507 and $\alpha_{s}(m_{Z})$ = 0.1298, 
respectively. The top pair production cross section of the SM prediction 
under the same condition is 197.6(1) pb.

\begin{figure}[htbp]
\begin{center}
\begin{tabular}{cc}
\includegraphics[width=8cm]{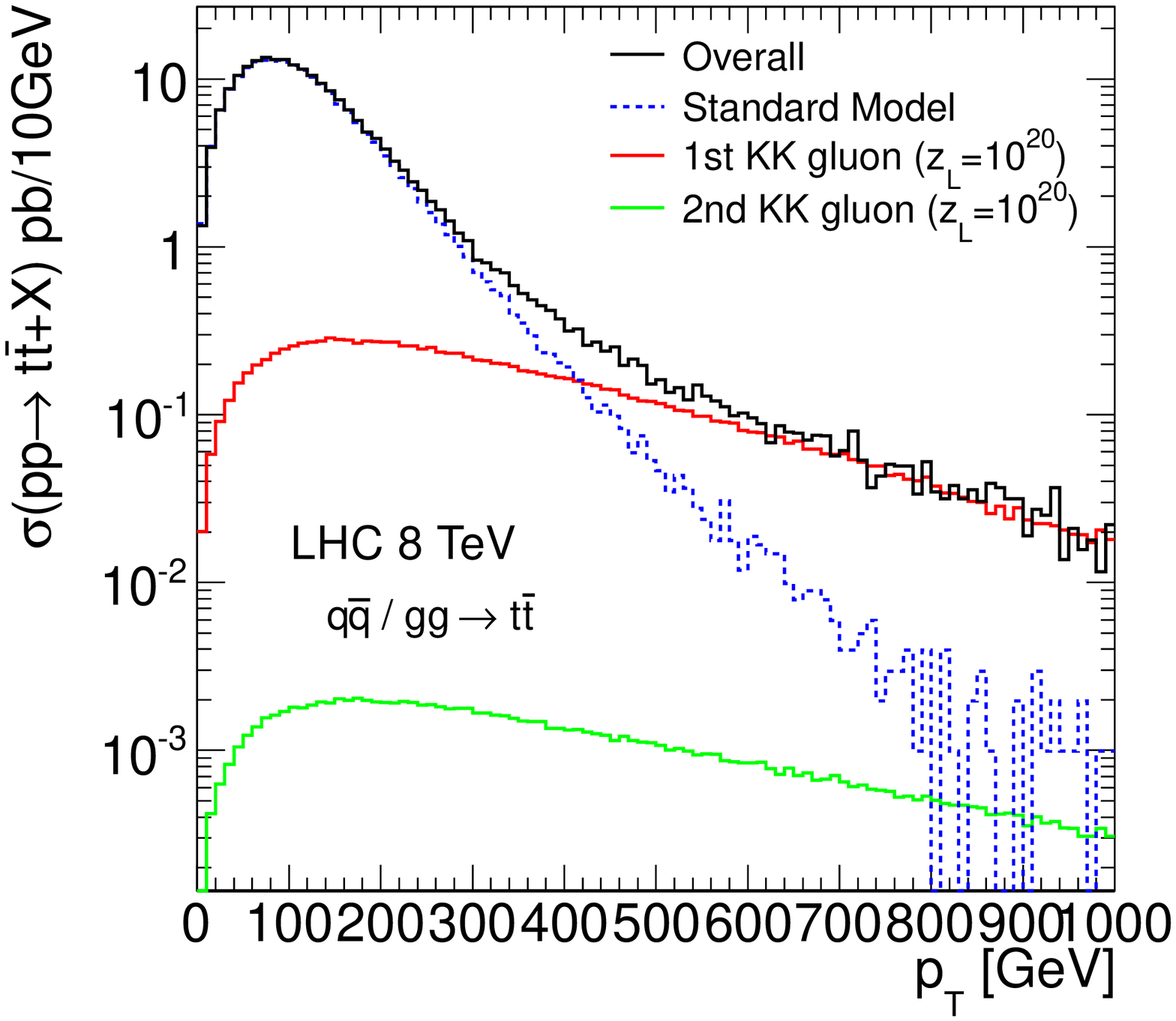} &
\includegraphics[width=8cm]{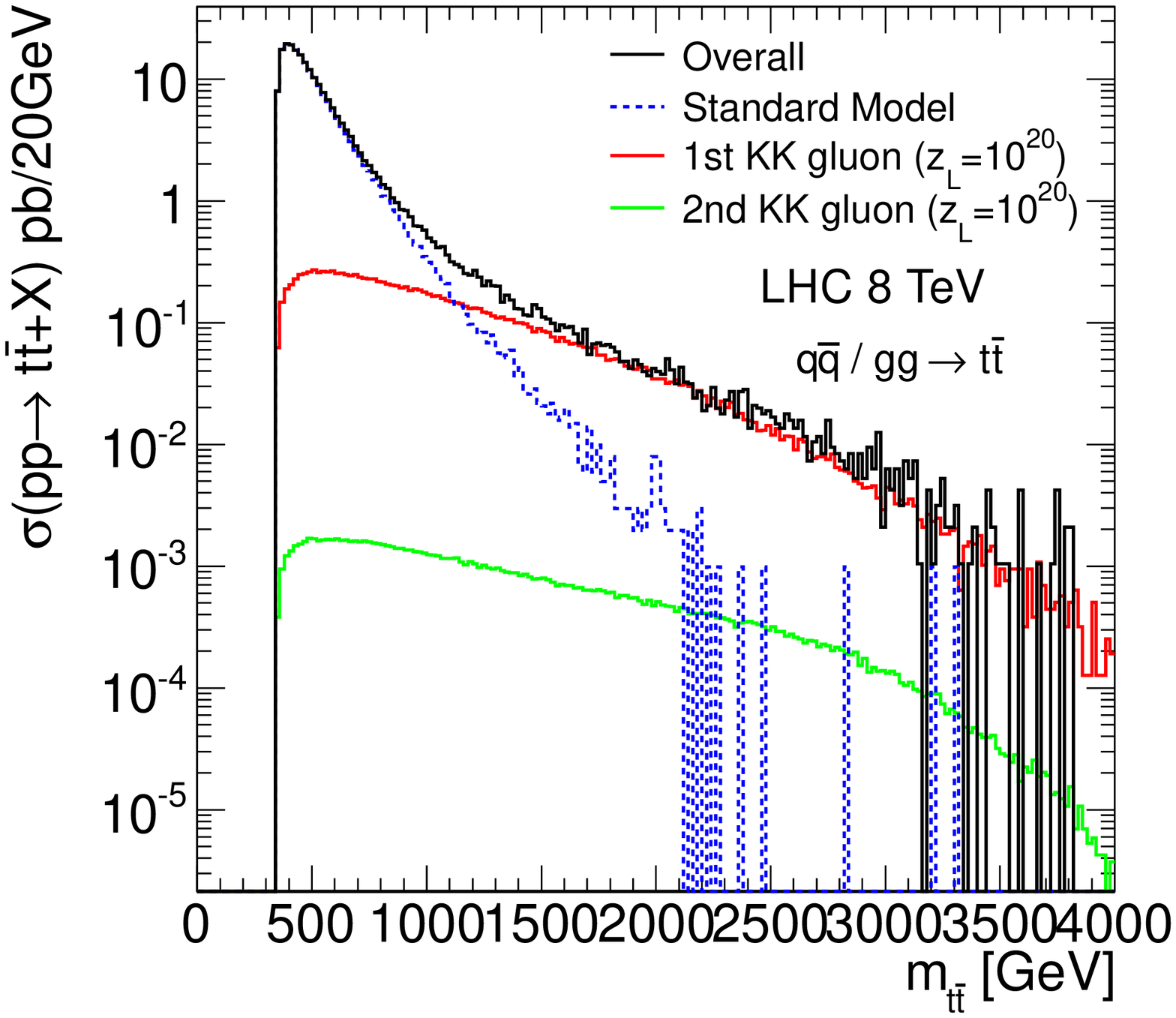} \\
(a) & (b) \\ \\
\end{tabular}
\caption{
(a) top quark $p_{T}$ spectrum and (b) $t\bar{t}$ invariant mass system $m_{t\bar{t}}$ with the first and second KK gluons
at $z_{L}$=10$^{20}$.
}
\label{topkinem}
\end{center}
\end{figure}

Figures \ref{topkinem} (a) and (b) present the top quark $p_{T}$ spectrum and the $t\bar{t}$ 
invariant mass system $m_{t\bar{t}}$ with the first and second KK gluons at 
$z_{L}$=10$^{20}$ together with the SM prediction, respectively. 
Both Figs. \ref{topkinem} (a) and (b) 
 show that the SM contribution is suppressed in high $p_{T}$ or $m_{t\bar{t}}$ regions,
and the first KK gluon production process becomes almost dominant. 
The top 
pair production cross section is precisely measured within 10\% level \cite{:2012xh}. Given the fact 
that the theoretical uncertainty also gives similar uncertainty at NNLO calculation \cite{Kidonakis:2012db}, 
the shown production cross sections in the table are nearly in the border of the experimental 
uncertainty. The differential cross section measurements \cite{Silva:2012di} as a function of top quark 
$p_{T}$ or $m_{t\bar{t}}$ will allow to explore a wide range of the parameter space of this 
model.

\begin{table}[tbhp]
\begin{center}
\begin{tabular}{c|ccc} \hline \hline
unit pb           & 1st KK gluon & 2nd KK gluon & 3rd KK gluon \\ \hline
$z_{L}$=10$^{15}$ & 22.61(2) & 0.1573(2) & 6.45(1)$\times$10$^{-6}$ \\
$z_{L}$=10$^{20}$ & 12.67(1) & 0.1065(1) & 8.50(1)$\times$10$^{-8}$ \\ \hline \hline
\end{tabular}
\end{center}
\caption{
Production cross sections of the first, second and third KK-gluons 
with the parameters of $z_{L}$=$10^{15}$ and $10^{20}$ in a unit of pb. 
The top quark mass is 172.5 GeV and CTEQ6L PDF is used for proton-proton collision at 
$\sqrt{s}$ = 8 TeV. The Standard Model top pair production cross section is 197.6(1) pb
under same condition.
}
\label{topxsec}
\end{table}

\subsection{Asymmetry measurement}

\begin{figure}[htbp]
\begin{center}
\includegraphics[width=8cm]{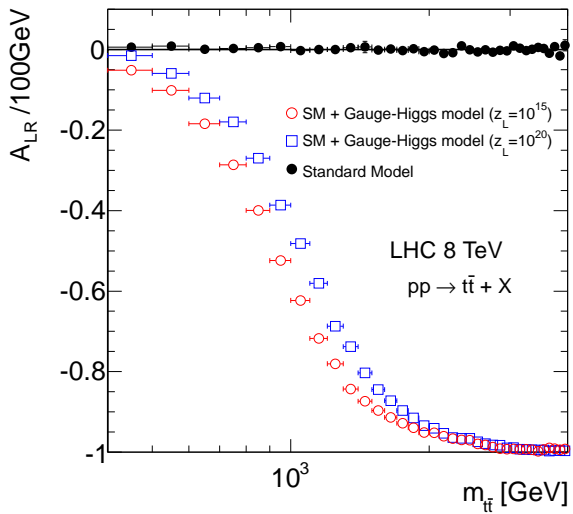}
\caption{
Left-right asymmetry, $A_{LR}$,
 as a function of the $t\bar{t}$ invariant mass system.
}
\label{topasymlr}
\end{center}
\end{figure}

\begin{figure}[htbp]
\begin{center}
\begin{tabular}{cc}
\includegraphics[width=8cm]{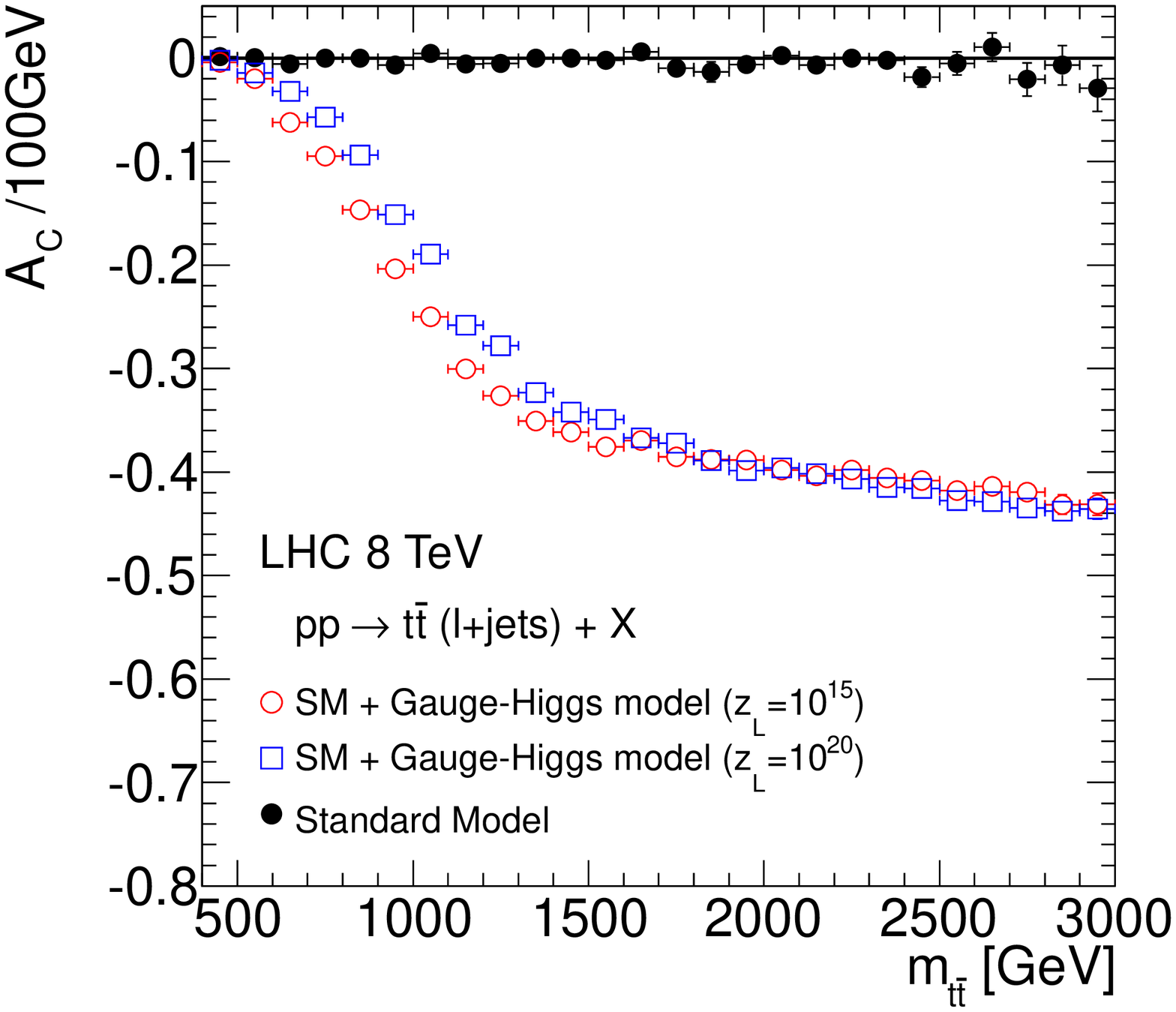} &
\includegraphics[width=8cm]{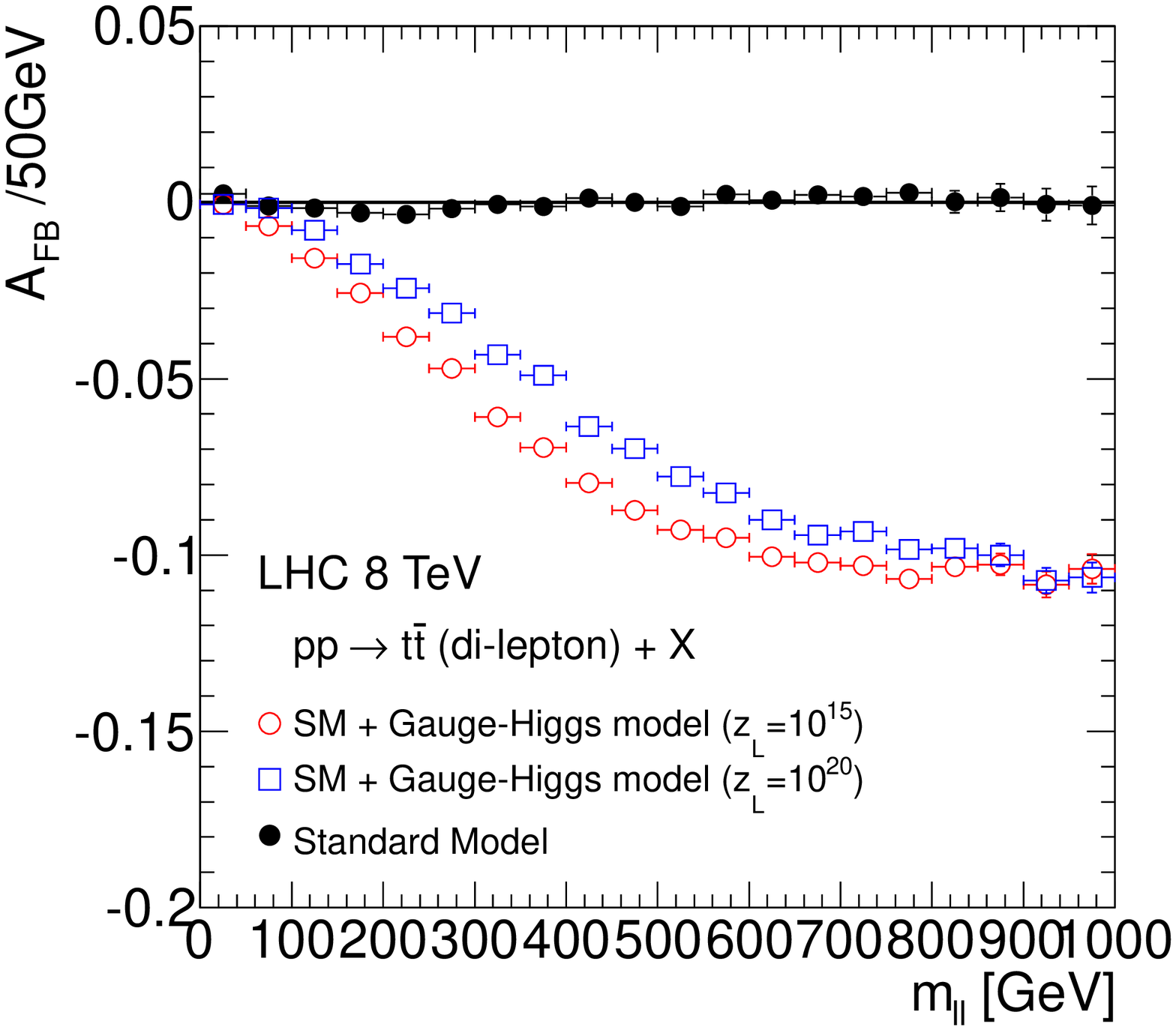} \\
(a) & (b) \\ \\
\end{tabular}
\caption{
(a) charge asymmetry and (b) forward-backward asymmetry as a function of the invariant mass of the 
$t\bar{t}$ system,
 after event selections applied for
 lepton + jets and di-lepton channels, respectively.
}
\label{topasym}
\end{center}
\end{figure}

Now let us estimate QCD parity violation 
 in the gauge-Higgs unification model. 
The left-right asymmetry is given as
\begin{equation}
  A_{LR} \; = \; \frac{N(t_{L}\bar{t}_{L}) - N(t_{R}\bar{t}_{R})}{N(t_{L}\bar{t}_{L}) + N(t_{R}\bar{t}_{R})}
\end{equation}
where $N$ is the number of events with left- or right-handed helicity state of the $t$ ($t_{L/R}$) and 
$\bar t$ ($\bar{t}_{L/R}$) quarks. First we present the left-right asymmetry $A_{LR}$ as a function 
of the $t\bar{t}$ invariant mass system in
 Fig. \ref{topasymlr}. The first, second and third KK gluons 
are interfered with the SM processes.
In the figure, the parameters $z_{L}$=10$^{15}$ and 10$^{20}$ are taken 
 in intuitive purpose. There is a very strong correlation in the
 asymmetric behavior of the left-
 and right-handed 
 helicity states of the $t\bar{t}$ production in the gauge-Higgs unification model, while there is no asymmetric behavior in the SM 
prediction. This is expected that the KK gluons are strongly coupled with the right-handed top quark.
This asymmetric behavior in
 Fig. \ref{topasymlr} can be quantitatively understood as follows.
In the high energy limit, $A_{LR}$ becomes 
\begin{align}
  A_{LR}
&\sim \frac{(g^{G^{(n)}}_{t_L})^2-(g^{G^{(n)}}_{t_R})^2}{(g^{G^{(n)}}_{t_L})^2+(g^{G^{(n)}}_{t_R})^2}.
\end{align}
Thus,  
$A_{LR}$ is close to $-1$ because $g^{G^{(n)}}_{t_L}$ is much smaller than $g^{G^{(n)}}_{t_R}$.
Even with the higher order correction,
 the SM only predicts at most less than 2\% 
\cite{Beenakker:1993yr}. 
Therefore, the size of 
 the asymmetry might be sufficient to observe in the experiment. 
Notice again that the helicity state
 is not identified 
 in the hadron collider experiments,
 because the $t$ ($\bar t$) quark is not a direct observable.
The $t$ ($\bar t$)
 quark is immediately decayed into the
 final state particles without suffering the strong interaction, 
 so that the correlation of the helicity state in the $t\bar{t}$ production is only known through the observation 
of the final state particle.
Since we can not directly know if $t$ ($\bar t$)
 is left- or right-handed,
 we should consider the asymmetry of the
 final state particles ($t\rightarrow bq\bar{q}$/$bl\nu$).
In order to observe the asymmetry in the experiment,
 we simulate an event 
 close to the experimental conditions. 
In the rest of this section,
 we calculate $t\bar t$ asymmetry as charge asymmetry
 and forward-backward asymmetry
 defined in Eqs.(3.6) and (3.7), and results are shown
 in Fig. \ref{topasym}.

The Matrix Element\cite{Tsuno:2006cu} of the top pair productions is considered 
up to the final state particles involving a decay of the $t$ ($\bar{t}$) quark, 
 so that the 
 helicity state in the $t$ ($\bar t$) quark
 in production is properly propagated into the final state 
 particles, and thus the event kinematics could be experimentally modeled. 
For simplicity, 
the event selections listed in Table \ref{ES} are 
applied based on the experimental signatures,
where the events are categorized as ``lepton + jets'' and ``di-lepton'' channels based on the $W$ boson 
decay from the $t$ and $\bar t$ quarks. 
%
\begin{table}[tbhp]
\begin{center}
\begin{tabular}{c|c} \hline \hline
                           channel &  event selection \\ \hline
lepton $+$ jets channel &  $p_T > 20$ GeV, $|\eta| < 2.5$ for lepton and quarks\\ \hline
\multirow{2}{*}{di-lepton channel} &  $p_T > 20$ GeV, $|\eta| < 2.5$ for leptons and quarks\\
                                           &  $\sqrt{\displaystyle{\sum_{\nu,\bar\nu}} p_x^2 + \displaystyle{\sum_{\nu,\bar\nu}} p_y^2} > 50$ GeV for neutrinos\\ \hline \hline
\end{tabular}
\end{center}
\caption{
Event selections, which categorized as ``lepton $+$ jets" and ``di-lepton" channels based on the $W$ boson decay from the $t$ and $\bar t$ quarks.
}
\label{ES}
\end{table}
The lepton + jets channel requires at least one high $p_{T}$ electron 
or muon in the fiducial volume in the detector. The $|\eta|<2.5$ is chosen by the coverage of typical 
tracking detectors. The $b$-quark and the other quarks from $W$ boson decay are considered as a jet which 
has to be $p_{T}$ larger than 20 GeV with in $|\eta|<2.5$. The $b$-jet tagging might enhance the top pair 
events against background processes. In the di-lepton channel, two leptons ($e$ or $\mu$) are required in the 
final state. To further suppress the SM background processes, the missing transverse energy,
which is the vectored summation of two neutrino momenta in the transverse plane, is required to be larger than
50 GeV. In the experiment,
 the un-folding procedure is applied to the observed experimental quantities,
 and here we only evaluate the 4-vector level event
 topology to see if given event selections are still 
 feasible to observe the $t\bar{t}$ asymmetry to probe this model.

Based on these event selections, we define the quantities of the $t\bar{t}$ asymmetry as follows,
\begin{equation}
  A_{C} \; = \; \frac{N(\Delta |y| > 0) - N(\Delta |y| < 0)}{N(\Delta |y| > 0) + N(\Delta |y| < 0)},
  \quad \Delta |y| \equiv |y_{t}|-|y_{\bar t}|
\end{equation}
for lepton + jets channel, and 
\begin{equation}
  A_{FB} \; = \; \frac{|\cos\theta_{lep}^{+}| - |\cos\theta_{lep}^{-}|}{|\cos\theta_{lep}^{+}| + |\cos\theta_{lep}^{-}|} \label{Afb}
\end{equation}
for di-lepton channel, respectively. The $A_{C}$ is known as the charge asymmetry and $A_{FB}$ is the forward-backward 
asymmetry. 
The $A_{C}$ is the difference of the events with 
the $t$ $(\bar t)$ quark rapidities, which is parametrized by $\Delta |y|$.
The $t$ $(\bar t)$ quark direction is reconstructed by three jets from the $t$ $(\bar t)$ quark decay. 
In the di-lepton channel,
$t$ $(\bar t)$ quark momentum can not be determined.
Meanwhile, it is easy to see the charged lepton momentum, and
charged leptons in the final state are expected to maintain the asymmetry of $t$ and $\bar t$ direction.
Note that $A_{FB}$ of (\ref{Afb}) is not the same observable at the Tevatron
because an absolute value of $t$ $(\bar t)$ flight direction is meaningful at the LHC.
$A_{FB}$ is formed by the event-by-event basis with the positive and negative 
charged 
lepton direction.

Figure \ref{topasym} (a) shows the charge asymmetry 
 and (b) does 
 forward-backward asymmetry as a function of 
 the invariant mass of the $t\bar{t}$ system, 
 and di-lepton mass system after event selections applied for lepton + jets 
 and di-lepton channels, respectively. 
We demonstrate the asymmetries when the gauge-Higgs unification model is 
included in the SM processes.
We also estimate that the integrated $A_C$ is $-0.04$.
As for $A_{FB}$, it reaches $-0.1$ in high $m_{ll}$ region.
With 5-10\% asymmetry, this is experimentally still sufficient to observe \cite{Silva:2012di}. 

\section{Conclusion}

We have 
discussed
 the parity violation in QCD 
 in warped extra dimension model
 where  
 Higgs and KK gluon are localized on the TeV brane 
 and left- and right-handed fermions have
 different configulations in the bulk. 
In this setup, 
 parity is violated 
in QCD sector at tree level, 
 which can be observed by the 
 helicity asymmetry of $t\bar t$ 
 at LHC. 
We pick up $SO(5)\times U(1)$ gauge-Higgs unification model 
 as a concrete model, 
 and find that large helicity asymmetry appears at
 high $m_{t\bar t}$ region.
We have evaluated LHC observable quantities,
 $A_C$ and $A_{FB}$, which can reach 
 $-0.4$ and $-0.1$, respectively,
 with specific parameters in high energy region.
We have 
 also estimated the integrated $A_C$ at $-0.04$ for all $m_{t\bar t}$ region.
Clearly, it is larger than the SM background,
and it is sufficient to be experimentally observed.
Furthermore, note that the threshold behavior of the asymmetry in the $t\bar{t}$ invariant mass system is 
proportional to the first KK gluon mass, and also the saturation behavior of the asymmetry is sensitive to 
the $z_{L}$ parameter. 

In the experimental side,
$A_{FB}$ is not yet reported at the LHC.
However, 
it is important because di-lepton channel is expected to be a probe of the $t\bar{t}$ asymmetry as we discussed in Section 3.2.
As for $A_C$ measurement, statistics is not sufficient in $m_{t\bar t}>450$ GeV region.
Thus, in this region, only an integrated $A_C$ is reported.
The measured charge asymmetry $A_C$ is consistent with our calculation \cite{Chatrchyan:2011hk}, 
however, total error (statistic and systematic) is still large ($\sim$ 5\%).
Statistic and systematic errors are approximately $\pm 0.03$(stat) and $\pm 0.02$(syst), respectively.
It is necessary to distinguish the KK gluon contribution from the SM background.
The SM prediction of $A_C$ is given by $A_C^{SM} = 0.00115 \pm 0.0006$\cite{Kuhn:2011ri}.
Then we can recognize the asymmetry to be KK gluon contribution when errors are suppressed as
$\sim 0.01$.
In oder to obtain $\sim 0.01$ statistic error,
the integrated luminosity need to be $\sim 100 ~{\rm fb}^{-1}$.
While, if the systematic error reduces about 1/10, 
it is possible to distinguish the $A_C$ evaluated in this paper from the SM background.
Therefore, 
$A_C$ is a promising observable for the KK gluon contribution. 
$A_{FB}$ is also hopeful observable in di-lepton channel, and
the precise measurements of 
$A_C$ and $A_{FB}$
at LHC are important to determine the 
coupling structure of this model.

\section*{Acknowledgments}

We thank Y. Hosotani for helpful discussions.
This work is partially supported by Scientific Grant by Ministry of 
 Education and Science, Nos. 22011005, 24540272, 20244028, and 21244036.
The works of K.K. are supported by Research Fellowships of the Japan Society for the Promotion of Science for Young Scientists.


\begin{thebibliography}{99}

\bibitem{CMS-ATLAS}
G.~Aad {\it et al.}  [ATLAS Collaboration],
  Phys.\ Lett.\ B {\bf 716} (2012) 1
  [arXiv:1207.7214 [hep-ex]];
  S.~Chatrchyan {\it et al.}  [CMS Collaboration],
  Phys.\ Lett.\ B {\bf 716} (2012) 30
  [arXiv:1207.7235 [hep-ex]].

  

\bibitem{Cabibbo:1979ay}
  N.~Cabibbo, L.~Maiani, G.~Parisi and R.~Petronzio,
  Nucl.\ Phys.\ B {\bf 158}, 295 (1979).

\bibitem{EliasMiro:2011aa}
  J.~Elias-Miro, J.~R.~Espinosa, G.~F.~Giudice, G.~Isidori, A.~Riotto and A.~Strumia,
  Phys.\ Lett.\ B {\bf 709}, 222 (2012)
  [arXiv:1112.3022 [hep-ph]].


\bibitem{Djouadi:2005gi} 
  A.~Djouadi,
  Phys.\ Rept.\  {\bf 457}, 1 (2008)
  [hep-ph/0503172].


%
%
  
  
\bibitem{Randall:1999ee}
  L.~Randall, R.~Sundrum,
  Phys.\ Rev.\ Lett.\  {\bf 83 } (1999)  3370-3373.
  [hep-ph/9905221].
  
\bibitem{Grossman:1999ra}
  Y.~Grossman and M.~Neubert,
  Phys.\ Lett.\ B {\bf 474} (2000) 361
  [hep-ph/9912408]; 
  T.~Gherghetta and A.~Pomarol,
  Nucl.\ Phys.\ B {\bf 586} (2000) 141
  [hep-ph/0003129]; 
  S.~Chang, C.~S.~Kim and M.~Yamaguchi,
  Phys.\ Rev.\ D {\bf 73} (2006) 033002
  [hep-ph/0511099];
  M.~E.~Albrecht, M.~Blanke, A.~J.~Buras, B.~Duling and K.~Gemmler,
  JHEP {\bf 0909} (2009) 064
  [arXiv:0903.2415 [hep-ph]].
  
  
\bibitem{Huber:2003tu}
  S.~J.~Huber,
  Nucl.\ Phys.\ B {\bf 666} (2003) 269
  [hep-ph/0303183];
  K.~Agashe, G.~Perez and A.~Soni,
  Phys.\ Rev.\ D {\bf 71} (2005) 016002
  [hep-ph/0408134];
  M.~S.~Carena, A.~Delgado, E.~Ponton, T.~M.~P.~Tait and C.~E.~M.~Wagner,
  Phys.\ Rev.\ D {\bf 71} (2005) 015010
  [hep-ph/0410344];
  E.~De Pree and M.~Sher,
  Phys.\ Rev.\ D {\bf 73} (2006) 095006
  [hep-ph/0603105];
  M.~S.~Carena, E.~Ponton, J.~Santiago and C.~E.~M.~Wagner,
  Nucl.\ Phys.\ B {\bf 759} (2006) 202
  [hep-ph/0607106];
  G.~Cacciapaglia, C.~Csaki, J.~Galloway, G.~Marandella, J.~Terning and A.~Weiler,
  JHEP {\bf 0804} (2008) 006
  [arXiv:0709.1714 [hep-ph]];
  C.~Csaki, A.~Falkowski and A.~Weiler,
  Phys.\ Rev.\ D {\bf 80} (2009) 016001
  [arXiv:0806.3757 [hep-ph]];
  S.~Casagrande, F.~Goertz, U.~Haisch, M.~Neubert and T.~Pfoh,
  JHEP {\bf 0810} (2008) 094
  [arXiv:0807.4937 [hep-ph]];
  M.~E.~Albrecht, M.~Blanke, A.~J.~Buras, B.~Duling and K.~Gemmler,
  JHEP {\bf 0909} (2009) 064
  [arXiv:0903.2415 [hep-ph]];
  M.~Bauer, F.~Goertz, U.~Haisch, T.~Pfoh and S.~Westhoff,
  JHEP {\bf 1011} (2010) 039
  [arXiv:1008.0742 [hep-ph]].

  
  
\bibitem{Lillie:2007yh}
  B.~Lillie, L.~Randall and L.~-T.~Wang,
  JHEP {\bf 0709} (2007) 074
  [hep-ph/0701166];
  B.~Lillie, J.~Shu and T.~M.~P.~Tait,
  Phys.\ Rev.\ D {\bf 76} (2007) 115016
  [arXiv:0706.3960 [hep-ph]]; 
  A.~Djouadi, G.~Moreau and R.~K.~Singh,
  Nucl.\ Phys.\ B {\bf 797} (2008) 1
  [arXiv:0706.4191 [hep-ph]].
  
  
%
\bibitem{Stelzer:1995gc}
  T.~Stelzer, S.~Willenbrock,
  Phys.\ Lett.\  {\bf B374 } (1996)  169-172.
  [hep-ph/9512292];
  %
  M.~Beneke, I.~Efthymiopoulos, M.~L.~Mangano, J.~Womersley, A.~Ahmadov, G.~Azuelos, U.~Baur, A.~Belyaev {\it et al.},
  [hep-ph/0003033];
%
  W.~Bernreuther, A.~Brandenburg, Z.~G.~Si and P.~Uwer,
  Phys.\ Rev.\ Lett.\  {\bf 87} (2001) 242002
  [arXiv:hep-ph/0107086];
  W.~Bernreuther, A.~Brandenburg, Z.~G.~Si and P.~Uwer,
  Nucl.\ Phys.\  B {\bf 690} (2004) 81
  [arXiv:hep-ph/0403035];
  W.~Bernreuther, M.~Fucker, Z.~-G.~Si,
  Phys.\ Rev.\  {\bf D74 } (2006)  113005.
  [hep-ph/0610334].
  W.~Bernreuther,
  J.\ Phys.\ G {\bf G35 } (2008)  083001.
  [arXiv:0805.1333 [hep-ph]].


\bibitem{Beenakker:1993yr}
  W.~Beenakker, A.~Denner, W.~Hollik, R.~Mertig, T.~Sack, D.~Wackeroth,
  Nucl.\ Phys.\  {\bf B411 } (1994)  343-380;
  C.~Kao and D.~Wackeroth,
  Phys.\ Rev.\  D {\bf 61} (2000) 055009
  [arXiv:hep-ph/9902202];
  W.~Bernreuther, M.~Fucker and Z.~G.~Si,
  Phys.\ Rev.\  D {\bf 78} (2008) 017503
  [arXiv:0804.1237 [hep-ph]].
  
\bibitem{Haba:2011cj}
  N.~Haba, K.~Kaneta, S.~Matsumoto, T.~Nabeshima and S.~Tsuno,
  Phys.\ Rev.\ D {\bf 85} (2012) 014007
  [arXiv:1109.5082 [hep-ph]].
\bibitem{Haba:2011ib}
  N.~Haba, K.~Kaneta and T.~Onogi,
  arXiv:1109.5442 [hep-ph].
  
 
%
%
%
%
%
  


  
  
 
\bibitem{Hosotani:2011vr}
  Y.~Hosotani, M.~Tanaka and N.~Uekusa,
  Phys.\ Rev.\ D {\bf 84} (2011) 075014
  [arXiv:1103.6076 [hep-ph]].
  
  
  



  
\bibitem{Blanke:2008zb}
  M.~Blanke, A.~J.~Buras, B.~Duling, S.~Gori and A.~Weiler,
  JHEP {\bf 0903} (2009) 001
  [arXiv:0809.1073 [hep-ph]];
  A.~J.~Buras,
  PoS KAON {\bf 09} (2009) 045
  [arXiv:0909.3206 [hep-ph]].


  


  
  

  

%
\bibitem{Pumplin:2002vw}
  J.~Pumplin, D.~R.~Stump, J.~Huston, H.~L.~Lai, P.~M.~Nadolsky and W.~K.~Tung,
  JHEP {\bf 0207} (2002) 012
  [hep-ph/0201195].


\bibitem{:2012xh}
  V.~Khachatryan {\it et al.}  [CMS Collaboration],
  Phys.\ Lett.\ B {\bf 695} (2011) 424
  [arXiv:1010.5994 [hep-ex]];
  G.~Aad {\it et al.}  [ATLAS Collaboration],
  Phys.\ Lett.\ B {\bf 717} (2012) 89
  [arXiv:1205.2067 [hep-ex]].
  
  
  
  
\bibitem{Kidonakis:2012db}
  N.~Kidonakis,
  arXiv:1205.3453 [hep-ph].
\bibitem{Silva:2012di}
  P.~Silva {\it et al.}  [ATLAS and CMS Collaborations],
  arXiv:1206.2967 [hep-ex].

\bibitem{Tsuno:2006cu}
  S.~Tsuno, T.~Kaneko, Y.~Kurihara, S.~Odaka and K.~Kato,
  Comput.\ Phys.\ Commun.\  {\bf 175} (2006) 665
  [hep-ph/0602213].
  
\bibitem{Chatrchyan:2011hk}
  S.~Chatrchyan {\it et al.}  [CMS Collaboration],
  Phys.\ Lett.\ B {\bf 717} (2012) 129
  [arXiv:1207.0065 [hep-ex]];
  G.~Aad {\it et al.}  [ATLAS Collaboration],
  Eur.\ Phys.\ J.\ C {\bf 72} (2012) 2039
  [arXiv:1203.4211 [hep-ex]].
  
\bibitem{Kuhn:2011ri}
  J.~H.~Kuhn and G.~Rodrigo,
  JHEP {\bf 1201} (2012) 063
  [arXiv:1109.6830 [hep-ph]].
\end{thebibliography}
\end{document}